# The Capture of Trojan Asteroids by the Giant Planets During Planetary Migration

P. S. Lykawka[1*#] and J. Horner[2†]

[1] International Center for Human Sciences (Planetary Sciences), Kinki University, 3-4-1 Kowakae, Higashiosaka-shi, Osaka-fu, 577-8502, Japan

[2] Dept. of Physics and Astronomy, The Open University, Walton Hall, Milton Keynes, MK7 6AA, UK

[*] E-mail address: patryksan@gmail.com
[#] PSL was earlier with Kobe University, Aepartment of Earth and Planetary Sciences, Kobe, Japan.
[†] JH is now at the Department of Physics, Durham University, Science Laboratories, South Road, Durham, UK, DH1 3LE

**ABSTRACT**

Of the four giant planets in the Solar system, only Jupiter and Neptune are currently known to possess swarms of Trojan asteroids – small objects that experience a 1:1 mean motion resonance with their host planet. In Lykawka et al. (2009), we performed extensive dynamical simulations, including planetary migration, to investigate the origin of the Neptunian Trojan population. Utilising the vast amount of simulation data obtained for that work, together with fresh results from new simulations, we here investigate the dynamical capture of Trojans by all four giant planets from a primordial trans-Neptunian disk. We find the likelihood of a given planetesimal from this region being captured onto an orbit within Jupiter's Trojan cloud lies between several times $10^{-6}$ and $10^{-5}$. For Saturn, the probability is found to be in the range $<10^{-6}$ to $10^{-5}$, whilst for Uranus the probabilities range between $10^{-5}$ and $10^{-4}$. Finally, Neptune displays the greatest probability of Trojan capture, with values ranging between $10^{-4}$ and $10^{-3}$. Our results suggest that all four giant planets are able to capture and retain a significant population of Trojan objects from the disk by the end of planetary migration. As a result of encounters with the giant planets prior to Trojan capture, these objects tend to be captured on orbits that are spread over a wide range of orbital eccentricities and inclinations. The bulk of captured objects are to some extent dynamically unstable, and therefore the populations of these objects tend to decay over the age of the Solar System, providing an important ongoing source of new objects moving on dynamically unstable orbits among the giant planets. Given that a huge population of objects would be displaced by Neptune's outward migration (with a potential cumulative mass a number of times that of the Earth), we conclude that the surviving remnant of the Trojans captured during the migration of the outer planets might be sufficient to explain the currently known Trojan populations in the outer Solar system.

# 1 INTRODUCTION

Two of the four giant planets in our Solar system, Jupiter and Neptune, are attended by swarms of objects known as Trojans, small bodies that move approximately 60º ahead and behind the planet in its orbit (known as the L4 and L5 Lagrange points). Such orbits, upon which bodies librate around the location of one or other of these Lagrange points, are often called "tadpole orbits" (Chebotarev 1974). Objects moving on typical Trojan orbits are protected from close encounters with their controlling planet by the effects of the 1:1 mean motion resonance (MMR) of that planet. This enables such objects to remain in the Trojan clouds on Gyr timescales. The structure of these swarms bears the imprint of events that happened during the formation of our Solar system, and models of the processes that occurred during that formation period must take great pains to both explain the currently known populations of Trojan objects, and make predictions of future discoveries that will allow those models to be tested critically (Fleming & Hamilton 2000; Morbidelli et al. 2005; Lykawka & Mukai 2008).

In a previous work (Lykawka et al. 2009), we examined the evolution of populations of objects that either formed within Neptune's Trojan clouds, or were captured to them, as the planet migrated outwards following one of four different migration scenarios through the outer Solar system. Here, we return to the vast amount of data obtained in those simulations in an attempt to answer a slightly different question – did the outward migration of Neptune (and the contemporaneous movement of the other giant planets) cause objects that formed beyond that planet to be captured within the Trojan clouds of the other giant planets?

It is taken as fact that, at the current epoch, the dynamics of the outer Solar system are such that the L4 and L5 Lagrange points in the orbits of Saturn and Uranus are unable to house a substantial population of stable Trojans, whilst the Trojan clouds of Jupiter and Neptune contain large populations that exhibit dynamical stability on the Gyr timescale (Nesvorny & Dones 2002). The instability of the Saturnian and Uranian Trojan clouds has one surprising effect, however – since the regions are easier to escape, they are also easier to enter (dynamics being a time-reversible process), and so temporary captures of objects within these regions can be expected to be intrinsically more common for these planets than for Jupiter or Neptune (e.g. Horner & Evans 2006). Whether the long-term instability of these regions is justified (something we hope to reconfirm in future work), this clearly poses a secondary question – did Saturn and Uranus acquire significantly larger captured Trojan populations than Jupiter and Neptune, during their migration? If this were the case, given that the current populations of the Jovian and Neptunian clouds are believed to greatly exceed that of the main belt of asteroids (Sheppard & Trujillo 2006), what would be the effect on the impact regime in the entire Solar system of the post-migration decay of these captured clouds? It seems possible that such release could extend any period of enhanced impact flux resulting from planetary migration on the terrestrial planets, which may go some way to explaining the postulated Late Heavy Bombardment of these worlds (Chapman, Cohen & Grinspoon 2007 and references therein).

In this work, we take advantage of detailed dynamical simulations in order to better ascertain the amount of material that would be acquired as Trojans of the giant planets as a result of their migration through the outer Solar system. In Section 2, we briefly describe the technique used in this modelling (more details of which can be found in Lykawka et al. 2009), before detailing our results in Section 3. In Section 4, we discuss the nature of the mechanism through which these Trojans are captured, before finally, in the last two sections, presenting a comprehensive discussion of our main results, our conclusions, and the avenues available for future work.

# 2 MODELLING

Recent models that aim to explain the dynamical evolution of the giant planets and the orbital architecture of the Edgeworth-Kuiper and asteroid belts require that these planets underwent significant radial displacement (planetary migration) during the early Solar system (Hahn & Malhotra 2005; Levison et al. 2008; Lykawka & Mukai 2008 and references therein). Following this paradigm, detailed dynamical simulations of the migration of the outer planets were performed using the n-body integrator *EVORB* (Brunini & Melita 2002). The migration of the planets was modelled using the following equation

$$a_k(t) = a_k(F) - \delta a_k \exp(-t/\tau) \qquad \text{Eq. 1}$$

where $a_k(t)$ is the semi-major axis of the planet after time $t$, $a_k(F)$ is the final (current) value of the semi-major axis, and $\tau$ is a constant determining the rate of migration of the planet. The fast and slow migration runs described above employed $\tau$ values of one and ten Myr, respectively, and the objects were followed for a period of $5\tau$ in both cases, after which the planets had reached their current locations. The index $k$ refers to the four giant planets, Jupiter ($k$ = J), Saturn ($k$ = S), Uranus ($k$ = U) and Neptune ($k$ = N). Such migration has been modelled in several previous studies (e.g. Malhotra 1995; Chiang et al. 2003; Hahn & Malhotra 2005), and represents a well accepted simplification for the migration process. We set the value of $\delta a_k$ so that the planets migrated from their starting locations to their current ones. Again, following these earlier works, Jupiter migrated inwards, while the other planets migrated outwards.

In each scenario considered, the planets Jupiter and Saturn had initial semi-major axes of 5.4 and 8.6 AU, respectively. Uranus and Neptune began the simulations at a variety of initial semi-major axes, as detailed in Table 1. In general, the orbital evolution of the giant planets was such that their eccentricities were ~0.04-0.05 for Jupiter and Saturn, and ~0.01-0.03 for Uranus and Neptune. A disk of $N_{disk}$ massless particles was placed on dynamically cold ($e < 0.01$, $i < 0.6°$) orbits located at semi-major axes between $a_{N0} + 1$ AU and 30 AU. The orbital evolution of these particles was followed until they collided with one of the giant planets, fell into the Sun, or experienced ejection from the Solar system (upon reaching a heliocentric distance of 200 AU). For a more detailed discussion of the simulations used, we refer the reader to Lykawka et al. (2009).

| Variant code | $a_{U0}$ (AU) | $a_{N0}$ (AU) | $\tau$ (Myr) | $N_{disk}$ |
|---|---|---|---|---|
| N18-F | 14.1 | 18.1 | 1 | 30,000 |
| N18-S | 14.6 | 18.1 | 10 | 100,000 |
| N23-F | 16.1 | 23.1 | 1 | 30,000 |
| N23-S | 16.2 | 23.1 | 10 | 80,000 |

**Table 1:** Details of the various simulations performed. $a_{U0}$ and $a_{N0}$ give the starting locations for Uranus and Neptune, while $\tau$ gives the migration scaling factor used (the planets migrated for a period of time equal to $5\tau$, after which they had reached their current locations). $N_{disk}$ gives the number of particles placed in the trans-Neptunian disk in each case.

**3 RESULTS**

We computed the number of particles that were captured as Trojans by Jupiter, Saturn, Uranus and Neptune at the end of planet migration for each of the models detailed in Table 1 through use of a modified version of *RESTICK* (Lykawka & Mukai 2007). The detection routines were optimised to enable the detection of any Trojan librating around either the L4 or L5 Lagrange point that had a Trojan lifetime greater than ~0.5 Myr (fast migration) or ~1 Myr (slow migration). These minimum Trojan lifetimes differ as a result of the shorter data output timescale which was used for the fast migration runs (a time step which would have been computationally prohibitive for the longer slow migration runs). Although our code was capable of detecting horseshoe Trojans for each of the giant planets, the resolution of the output data was such that it prevented us from accurately identifying all such objects, and avoiding false detections. Therefore, these bodies were not taken into account in the current investigation.

In order to estimate the efficiency with which the giant planets captured Trojan objects, we determined the number of objects that experienced at least one close encounter with each planet over the duration of the simulations. Any time a test particle approached a giant planet to within its Hill radius, it was considered to have experienced a close encounter with that planet.

In the following subsections, we discuss the results of our calculations for each giant planet in turn. It should be noted that, due to both the lengthy "minimum Trojan lifetime" required for an object to be detected as a Trojan, and the admittedly arbitrary definition of "close encounter", as described above, the capture efficiencies derived below are subject to uncertainties that, conservatively, should be considered to be a factor of a few. In addition, when compared with the capture efficiencies obtained for the case of fast migration, the true capture efficiencies for the slow migration runs should be somewhat larger than those obtained here, as a result of the larger "minimum lifetime" that was required for detection with *RESTICK*. However, we note that the obtained capture efficiencies presented in this work should be considered lower limits, because the lack of data output resolution prevented *RESTICK* from identifying the undoubtedly significant population of Trojans captured for periods shorter than the 0.5-1 Myr thresholds mentioned above. A discussion of the long-term stability of the captured Trojans can be found in Section 5.

### 3.1 THE CAPTURE OF TROJANS BY JUPITER

No objects were captured as Jovian Trojans in any of the four migration runs. This is likely, in part, due to the fact that very few particles in our simulations were able to evolve onto Jupiter-encountering orbits before the end of the integrations. Combined with this, an intrinsically low capture rate to the Jovian Trojan cloud means that no such objects were present in any of our runs. The lack of any captured objects, when combined with data on the total number of objects that evolved to encounter Jupiter across our simulations, allows us to place a coarse upper limit on the capture efficiency of Jovian Trojans (both L4 and L5 type orbits) of approximately $2\times10^{-4}$ (fast migration) and $2\text{-}3\times10^{-5}$ (slow migration).

The Jovian Trojan population has been known for over a hundred years, and several thousand of its members have already been discovered. As a result of this long observational history, a great number of studies have been carried out into this population, and, as such, it is clearly important to attempt to obtain a genuine value for the capture efficiency of objects to this population, and to test the upper limits determined above. In order to do this, we decided to perform an additional simulation of fast planetary migration using the same numerical procedures as described in Section 2, and including all four giant planets, but with a particular emphasis on the Jovian region[1]. Jupiter and Saturn started with eccentricities similar to those currently observed (~0.05). To allow us to better judge the capture efficiency of objects to the Jovian cloud, we greatly increased the number of test particles contained within the simulation. We therefore placed three million massless test particles on Jupiter-approaching orbits with initial semi-major axes randomly but uniformly distributed between 6 and 10 AU. These particles had inclinations in the range 0 – 30º, and perihelia which lay at q > 5.4 AU (i.e. able to approach, but not cross, Jupiter's initial orbit). All test particles had eccentricities > 0.1. This orbital element distribution was chosen to mimic that found for objects which evolved onto orbits close to the orbit of Jupiter from the trans-Neptunian disk after a few million years had elapsed in our initial runs (as described in Table 1), while ensuring that no Jovian Trojans existed at the start of the simulation. As the simulation progressed, the ratio between the orbital periods of Saturn and Jupiter ($P_S / P_J$) increased gradually from an initial value of ~2.045 to its current value of ~2.485, which was reached at the end of planetary migration. The number of Trojans resident in the Jovian cloud was calculated using *RESTICK* after five and ten million years of integration time, yielding totals of 149 (5 Myr) and 131 (10 Myr) objects. The captured objects cover wide ranges of eccentricity and inclination space ($e < 0.21$; $i < 45°$, see Fig. 1, top panel), as would be expected. The resulting capture efficiencies obtained from these results are approximately $5.0 \times 10^{-5}$ and $4.4 \times 10^{-5}$, respectively, a factor of ~4 smaller than the upper limit calculated from our initial results, as discussed above. Assuming that the capture efficiency in the slow migration case is similarly lower than the estimated upper limit would yield a capture efficiency on the order of several times $10^{-6}$.

---

[1] In this particular simulation, the four giant planets started at 5.4, 8.7, 16.1, and 23.1 AU, respectively. As described by the initial conditions of the disk particles, we focused only on Jupiter-encountering orbits, so that the initial settings for Uranus and Neptune were unimportant. Therefore, we regard this simulation as representative of fast migration on a general manner.

Although we have modelled the Solar system in a different manner, and a detailed comparison is therefore problematic[2], it is nevertheless interesting to compare our results with those obtained by Morbidelli et al. (2005) in their 'Nice model'. It seems that our initial conditions are relatively similar to the orbital architecture which occurs in that model shortly after Jupiter and Saturn have crossed their mutual 1:2 MMR. Indeed, the values we have obtained for the capture efficiency of Jupiter-encountering objects to that planet's Trojan cloud are in good agreement with those determined by Morbidelli et al. (2005), who obtained $1.8 \times 10^{-5}$ and $2.4 \times 10^{-6}$ for "fast" and "slow" migrations, respectively. We discuss the mechanism through which such capture events occur, and the implications of this result in more detail in Sections 4 and 5 of this work.

In addition to the extra-large simulation described above, we performed two further integrations, each studying the evolution of 500,000 test particles, in order to examine whether the initial orbital eccentricities of Jupiter and Saturn could have any effect on the capture efficiency of Trojans during their migration and the effects of a 3:7 MMR crossing event between Jupiter and Saturn[3]. The initial conditions for these extra simulations were the same as those described above, except that the initial eccentricities and inclinations of Jupiter and Saturn were set to be ~0.001, and their initial orbital radii were set such that $P_S / P_J \sim 2.3$ (so they began the integration close to a mutual 3:7 MMR crossing event), respectively. After 5 Myr had elapsed in the two integrations, we found that 34 and 16 objects had been captured as Jupiter Trojans, yielding capture efficiencies of $7 \times 10^{-5}$ and $3 \times 10^{-5}$, respectively. After 10 Myr, these numbers became $6 \times 10^{-5}$ and $2 \times 10^{-5}$.

Taken in concert with the results of the 3 million particle integration discussed above, it is clear that a variety of initial planetary architectures can yield fairly efficient capture of Trojans by Jupiter. When that planet begins migration from a fairly circular orbit just beyond the location of the 1:2 MMR with Saturn, capture rates of several $10^{-5}$ are observed, while capture during a migration involving only a late MMR crossing (such as the 3:7 MMR) still yields efficiencies of a few $10^{-5}$. Each tested scenario resulted in captured Trojans which displayed wide ranges of eccentricity and inclination, which suggests in turn that the nature of the Jovian Trojan cloud may not be tightly bound to the exact details of that planet's migration.

**3.2 THE CAPTURE OF TROJANS BY SATURN**
A single Saturn Trojan was found, resulting from the N23-F migration run. As in the case of Jupiter, we attribute the observed low capture efficiency to a combination of the small number of objects that acquired Saturn-encountering orbits over the course of the simulations and an intrinsically low probability of Trojan capture. Based upon this single capture, and the frequency with which objects in our integrations became Saturn-crossing, we determine that the capture efficiency for Saturn Trojans has an upper limit of approximately $1.5 \times 10^{-4}$ and $1.5-2 \times 10^{-5}$ for fast and slow migration, respectively. These values are similar to those obtained for Jupiter Trojans (Section 3.1).

---

[2] Unfortunately, the initial conditions given in Morbidelli et al. (2005) do not allow a detailed comparison here. In particular, the orbital distributions of disk particles and migration rates ($\tau$) are not given in that paper. In addition, only the evolution of the approximate ratio between Jupiter and Saturn's orbital periods is shown, making a more detailed comparison somewhat problematic.

[3] Apart from the mutual 1:2 MMR, Jupiter and Saturn may have experienced other important resonant crossings as well, such as the 3:7, 4:9, and 5:11, during planetary migration.

The detection of a solitary Saturnian Trojan at the end of our calculations shows that the capture of Trojans by Saturn is plausible. Given that many Earth-masses ($M_\oplus$) of material were likely scattered from the trans-Neptunian disk into the realm of the giant planets during Neptune's migration, this suggests that a substantial population of objects were captured as Saturnian Trojans over that period (see Section 5). These objects would be expected to display a wide range of orbital properties ($e, i$), since such excited orbital distributions are typically acquired by small bodies undergoing a protracted sequence of scattering events by the giant planets during planetary migration. The orbital evolution of the single captured particle is shown in Figure 2.

As in the case of the Jovian Trojans, extra simulations involving millions of test particles would be required in order to obtain both accurate capture efficiencies and detailed information on the other properties of captured Saturnian Trojans. However, since such simulations are computationally expensive and would require a wider range of initial conditions (to better describe the feeding mechanism), we prefer to leave this particular investigation for the future. It should be noted, however, that the theoretical upper limits calculated for the capture efficiency of Saturnian Trojans in this manner are almost equal to those calculated in Section 3.1 for the maximum capture efficiency of Jovian Trojans, prior to the execution of the additional simulations. Assuming that the capture efficiencies obtained for this planet would behave in the same manner as those obtained for the larger giant planet (in other words, that the true capture efficiency be equivalently smaller than the estimated upper limit), a conservative estimate of the capture efficiency for Saturnian Trojans can be assumed to be, at most, just a few times smaller than these upper limits, again yielding values of $<10^{-5}$ and $<10^{-6}$ for rapid and slow migration, respectively.

### 3.3 THE CAPTURE OF TROJANS BY URANUS
A significant number of Uranus Trojans were present at the end of three of our four main runs (specifically N18-F, N18-S and N23-F). This leads to capture efficiencies for the ice giant of approximately $5-6 \times 10^{-4}$ and $6 \times 10^{-5}$ for fast and slow migration[4], respectively. These values are several times higher than those found for Jupiter and Saturn, which can initially be taken to suggest that, during the period of planetary migration, Uranus was more likely to capture planetesimals as Trojans than either of those giant planets. This result can be interpreted in two ways. First, since the planetesimals were being scattered inward from beyond the orbit of Neptune, Uranus encountered a far greater number, over a longer period of its migration history, than either of the planets interior to it. This may have helped to increase the apparent capture efficiency over that recorded for Jupiter and Saturn. Second, since Uranus has a lower mass than either of those planets, it is less able to catastrophically alter the orbit of a given particle. This means that Uranus is less likely to scatter objects onto highly eccentric orbits, or to eject them completely from the Solar system, than its two larger siblings, which may lead to potential Trojans surviving in the vicinity of the planet for longer, giving them more opportunity to be captured in this way.

The Trojans captured by Uranus at the end of the migration period lay on orbits covering wide ranges of eccentricity and inclination ($0.05 < e < 0.3$; $i < 25°$, see Fig. 3). Unfortunately, too few objects were captured to allow any meaningful statistical comparison between the distributions of Uranian Trojan between the three migration scenarios which produced them. Nevertheless, the results show that the capture of disk objects into the Uranian Trojan clouds is a natural outcome of planetary migration, even when one only considers a dynamically cold planetesimal disk initially located beyond the orbit of Neptune.

---

[4] Value obtained for the N18-S run. The upper limit for the N23-S run was approximately $1 \times 10^{-5}$.

## 3.4 THE CAPTURE OF TROJANS BY NEPTUNE

Unsurprisingly, a relatively large number of Neptune Trojans were identified at the conclusion of planetary migration. The capture efficiency of objects to such orbits was found to lie in the range $2.7 – 12 \times 10^{-4}$ for fast migration, and the range $3.3 – 3.6 \times 10^{-4}$ for slow migration. These values are somewhat higher than those obtained for the capture of Trojans by Uranus in our runs.

The number of objects found to be Neptunian Trojans under the constraints described earlier in this section was nearly identical to that found (with tadpole orbits) from runs of significantly higher resolution that were detailed in Lykawka et al. (2009). This indicates that the modified version of *RESTICK* was well calibrated, and maintained its accuracy, yielding reliable results. These results are only included here for completeness, and to allow direct comparison with the other giant planets (Section 5). For a more detailed study of the capture and survival of objects in the Neptunian Trojan clouds, we direct the interested reader to our earlier work (Lykawka et al. 2009).

## 4 THE TROJAN CAPTURE MECHANISM

Recent studies strongly suggest that the currently observed Trojans of Jupiter and Neptune are the remnants of much larger populations that were captured in the distant past, during the process of planetary formation and migration (Morbidelli et al. 2005; Lykawka et al. 2009; Nesvorny & Vokrouhlicky 2009). In this work, we have shown that each of the giant planets should have captured significant Trojan populations by the end of planet migration. How does such capture occur? Here, we discuss the routes through which the giant planets captured Trojans during their migration.

Previous work by a number of authors has shown that disk planetesimals on planet-encountering orbits can experience enhanced temporary captures onto orbits within that planet's Trojan cloud after entering specific chaotic regions (Morbidelli et al. 2005; Horner & Evans 2006; Nesvorny & Vokrouhlicky 2009). In addition, Trojans can also escape back to the planetesimal disk under the same conditions (due to the time reversibility of the dynamics involved). Such chaotic regions originate from the overlap of secondary resonances associated with a mutual MMR between a pair of planets and the characteristic Trojan motion (the 1:1 MMR of an object with a given planet) (e.g. Murray & Dermott 1999).

Were the planets stationary with respect to one another, such regions of chaotic behaviour would remain fixed within the Solar system, and objects would be able to continually move in and out of the Trojan clouds. During the migration of the outer planets, however, it is clear that the 1:1 MMR of a given planet will pass through such chaotic regions. During that time, an increased flux of material will be able to flow both in and out of the Trojan clouds, until such a time that the 1:1 MMR moves away from such a chaotic regime. At this point, the displacement of the giant planets (as a direct result of their migration) breaks the time-reversibility of the situation in such a way that the orbits of previously transient captured Trojans can become "frozen" in – the objects becoming permanently captured. Once the chaotic regime has been left behind, both the capture of fresh Trojan material and the escape of old Trojans to the planetesimal disk effectively cease (e.g. Morbidelli et al. 2005). Such periods of enhanced Trojan capture/loss, followed by rapid "freeze-in" of the resulting population, can happen repeatedly as the planets migrate, as they may encounter several MMRs during their orbital evolution. By the end of planetary migration, dependant on the exact nature of that migration and the orbital properties of the disk population, a significant population of Trojans can be captured onto orbits that are stable on Gyr timescales.

Invoking the mechanism described above, Morbidelli et al. (2005) showed that the chaotic capture of Jovian Trojans is possible during discrete periods shortly after Jupiter and Saturn migrate through their mutual 1:2 MMR. The Jovian Trojans which were captured during our integrations were found to have experienced essentially the same capture. Indeed, as the planets migrated through the disk, the majority of captures happened between values of $P_S/P_J \sim$ 2.05 and 2.08 in our integrations, again confirming the results of Morbidelli et al. (2005). Additionally, we also found captures to be possible for a period of time when $P_S/P_J \sim$ 2.33 during the 3:7 MMR crossing, albeit at a somewhat lower efficiency.

Unfortunately, the solitary Saturnian Trojan capture reported in this work does not allow us to constrain the primary capture mechanism for this planet. Captures by Uranus and Neptune, however, were significantly more plentiful, which allowed us to determine the approximate capture times for those objects, and compare them to the times at which those two planets experienced mutual MMR crossings during their migration. In this manner, we were able to determine that approximately 2/3 of Uranian and 3/4 of Neptunian Trojans were captured during the first ~1.5τ of the simulations. As can be seen from Figure 4, this period spans the great bulk of the migration of these planets (thanks to the exponential nature of their motion), and they pass through all major mutual MMRs between their starting positions at $P_N/P_U \sim$ 1.45 and 1.71 (N18 and N23 runs, respectively) and their final locations at ~1.95 (as currently observed). In that time, the planets cross their mutual (2:3), (5:8), (3:5), 4:7, 5:9, 6:11 and 7:13 mutual MMRs (the resonances detailed in parentheses are only crossed for the N18 runs). The temporal distribution of captures to the Trojan clouds of these two planets shows significant "peaks" that match the timing of these MMR crossings, which strongly suggests that the main mechanism involved was again chaotic capture. These results confirm that Uranus-Neptune MMRs play an important role in the dynamics of the Neptune Trojans (Kortenkamp, Malhotra & Michtchenko 2004; Nesvorny & Vokrouhlicky 2009). It should, however, be noted that the capture times of the remaining 1/3 of Uranian and 1/4 of Neptunian Trojans do not match with these particular MMR crossings. Although a few of these objects were initially captured during this initial 1.5τ period, they then left the Trojan region and were re-captured at a later time. Other objects experienced dynamical histories that were significantly more complex, including periods of intermittent gravitational scattering by the giant planets and temporary captures in MMRs with Uranus or Neptune prior to entering the Trojan cloud. These cases may be explained by a similar chaotic capture mechanism, with capture resulting from chaotic regions arising from crossings of mutual MMRs between Uranus and Saturn and/or the contribution of the near 1:2 commensurability between Uranus and Neptune. In fact, it seems likely that the timing of the Uranus-Saturn MMR crossings reaching ~2.2 − 2.4τ of the integrations, and the approach of Uranus and Neptune to their mutual 1:2 MMR after ~4τ (Fig. 4) may explain the bulk of the late Trojan captures (>1.5τ) found in our runs.

## 5 DISCUSSION

In this work, we present the results of detailed dynamical simulations of the migration of the giant planets, and examine the fate of a debris disk stretching from just beyond the initial orbit of Neptune to its current location. As that giant planet migrates outwards through this disk, its constituent particles are scattered chaotically through the Solar system. A small but significant fraction of these objects are captured by Neptune and the other giant planets as Trojans, objects trapped in the 1:1 MMR of a given planet. The main route through which such objects become trapped appears to be chaotic capture. Such capture occurs when two of the giant planets undergo a mutual MMR crossing, which acts to destabilise the Trojan region of those planets, allowing the exchange of material between the Trojan clouds and the planetesimal disk. Once the planets move away from the MMR crossing, the Trojan population is "frozen in", as the Trojan cloud become significantly more stable. For any given scattered planetesimal the calculated capture probabilities are of the order of $10^{-4}$ (or lower) that that it will be captured as a Trojan for any planet. However, the simple fact that a huge population of objects would be displaced by Neptune's outward motion means that such low capture probabilities would result in each of the giant planets capturing large swarms of Trojans. One can certainly expect that, by the end of their migration, each giant planet was accompanied by a vast collection of captured Trojans, spread over a wide range of orbital eccentricities and inclinations, approximately 4 Gyr ago.

Our main and complementary simulations allowed us to determine that the likelihood of a given planetesimal being captured to the Jovian Trojan cloud lies in the range several times $10^{-6}$ to $10^{-5}$. Slightly smaller capture efficiencies can be expected for the capture of Trojans by Saturn, which suggests that this planet would have capture efficiencies on the order of $<10^{-6}–10^{-5}$. In the case of Uranus, the capture probability was found to be surprisingly high, approximately several times $10^{-5}–10^{-4}$, values which range from approximately an order of magnitude smaller than that for Neptune to a rate comparable to that planet. When followed over Gyr timescales, we obtained the following survival fractions for our captured populations of Jupiter, Uranus and Neptune Trojans: 25% (31 out of 131 bodies[5]; This work – See also Fig. 1, bottom panel), <4% (no survivor out of 25 bodies; This work) and a few percent, respectively (Horner & Lykawka 2009; Lykawka et al. 2009, in preparation).

How large (or massive) were the primordial captured Trojan populations in the outer Solar system? If we assume that the primordial planetesimal disk had had a surface density of material that followed a decaying power law of index -1.5, we can estimate the mass contained within the planetesimal disk. Assuming a conservative value of 1.5 gcm$^{-2}$ for the density of matter at 10 AU, the disks used in this work, stretching from 24-30 AU and 19-30 AU, would have contained approximately 13 and 25 $M_\oplus$ of material.

| Planet | $\varepsilon_{min}$ | $\varepsilon_{max}$ | $M_{min}$ ($M_\oplus$) | $M_{max}$ ($M_\oplus$) |
|---|---|---|---|---|
| Jupiter | $5\times10^{-6}$ | $5\times10^{-5}$ | $3\times10^{-5}$ | $2\times10^{-4}$ |
| Saturn | $<10^{-6}$ | $10^{-5}$ | $<8\times10^{-6}$ | $6\times10^{-5}$ |
| Uranus | $5\times10^{-5}$ | $5\times10^{-4}$ | $6\times10^{-4}$ | $7\times10^{-3}$ |
| Neptune | $3\times10^{-4}$ | $10^{-3}$ | $4\times10^{-3}$ | $2\times10^{-2}$ |

**Table 2:** Estimated minimum and maximum masses of the captured Trojan populations ($M_{min}$, $M_{max}$) for each of the giant planets at the end of their migration using approximate minimum and maximum capture efficiencies ($\varepsilon_{min}$, $\varepsilon_{max}$) (the values for Saturn are assumed, those for the other three giant planets are those obtained in this work). This takes into account the fraction of objects from the planetesimal disk observed to encounter the giant planets over the course of their migration (between 0.16 and 0.98, depending on which planet is considered, and the migration rate chosen), and assumes a mass of between 13 and 25 $M_\oplus$ of material was initially present in the disk.

Table 2 presents estimates of the minimum and maximum amount of mass that could have been captured as Trojans by the giant planets during their migration. It combines the fraction of objects from the primordial planetesimal disk which encounter the giant planets with minimum and maximum capture efficiencies as calculated in this work. It turns out that the fast the migration rate, or the larger the disk supplying planet-encountering objects, the greater the mass of objects captured as planetary Trojans. It should be noted that material sourced from other regions (such as the asteroid belt, and material from beyond the outer edge of the disk considered in this work) would doubtless add to the total mass captured by the giant planets, so, if anything, these values represent cautious lower limits to the total amount of material captured.

The fact that we have yet to observe Trojans of Saturn and Uranus suggests that these Trojans have been completely lost over the age of the Solar system. The current mass of material in the Jovian Trojan clouds is estimated to be at least of order $\sim10^{-5}$ $M_\oplus$, although some works suggest it may be in fact several times larger than that (Morbidelli et al. 2005; Fernandez, Jewitt & Ziffer 2009).

---

[5] For simplicity, only the results of the main special simulation were taken into account (Section 3.1). For completeness, we note that just 1 out of 11 Jovian Trojans captured during the 3:7 MMR crossing survived after 4 Gyr.

Comparing this value with our results above suggests that the lost Trojan populations of Saturn and Uranus were on the order of <1-6 and 60-700 times the mass of Jupiter's current population (which, we remind the reader, is thought to contain a comparable amount of material to the asteroid belt, at the current epoch), respectively! This also leads to another question: To what degree have the primordial populations of Jovian and Neptunian Trojans been dynamically depleted over the age of the Solar system? If we assume that the chaotic capture mechanism described in this work is the primary source of these Trojan populations, and consider only the largest Trojans (diameter, $D > 50$ km), in order that the effect of collisional grinding be negligible (e.g. Marzari et al. 1997), we can make use of earlier results to estimate the initial Jovian population based on the current mass. Since the great bulk of the mass contained within the Jovian Trojan population is housed by objects in this size range, these considerations seem reasonable. Given that we find that just 25% of captured Jovian Trojans would be expected to survive for 4 Gyr, we estimate the total primordial mass to have been $\sim 4 \times 10^{-5}$ $M_\oplus$. Levison et al. (2009) estimate that the survival fraction of Jovian Trojans to 4 Gyr could be as low as 13%, in which case our estimate of the initial post-migration mass of the Jovian Trojan cloud would grow to $\sim 8 \times 10^{-5}$ $M_\oplus$. A total expected mass in this range ($4 - 8 \times 10^{-5}$ $M_\oplus$) is in broad agreement with the value found through analysis of our standard simulations ($3 \times 10^{-5} - 2 \times 10^{-4}$ $M_\oplus$, determined above). Given that the estimates based on simulations are sensitive to the model parameters (such as disk mass and migration rate), while the estimates based on the current population are most sensitive to the estimated modern value of the total Jovian Trojan mass, it is interesting that the two values obtained are in such good agreement.

In the case of the Neptune Trojans, the situation is somewhat less certain, since the mass of the modern day population is still very poorly constrained. Estimates range as high as 20 times the mass of the Jovian population (Sheppard & Trujillo 2006). Assuming a somewhat more conservative mass of $\sim 10^{-4}$ $M_\oplus$, and taking into account that only a few percent of the captured Neptunian Trojans would survive after 4 Gyr, we estimate the total primordial mass of Neptune Trojans to be in the range $\sim 2 - 10 \times 10^{-3}$ $M_\oplus$ (assuming survival rates of 1% and 5%, respectively[6]), a value which is again in reasonable agreement with estimates taken from our standard simulations ($4 \times 10^{-3} - 2 \times 10^{-2}$ $M_\oplus$).

Consequently, taken together, the lost Trojans of Jupiter and Saturn probably contained 3-10 times the current mass of observed Jovian Trojans, which implies the release of $\sim 3 \times 10^{-5} - 10^{-4}$ $M_\oplus$ of material onto unstable orbits over the time since planetary migration ceased. On the other hand, the loss of Uranian and Neptunian Trojans probably amounted several tens or even hundreds times $10^{-5}$ $M_\oplus$, thus providing an important additional source of material on unstable orbits among the giant planets. Such unstable wanderers are known as the Centaurs, and represent the direct parent population of the Jupiter family of comets. This finding strengthens the idea that planetary Trojans have acted as a significant source of the Centaurs and related populations over the age of the Solar system, as proposed in Horner & Lykawka (2009). In addition, this dramatic loss of the entire population of Saturnian and Uranian Trojans (through dynamical decay), coupled with the escape of a substantial fraction of primordial Jovian and Neptunian Trojans, may have led to an enhanced impact flux of such lost Trojan objects on the host planets and their satellites.

Finally, given that the standard models of Trojan formation are unable to explain the highly dynamically excited orbits of a large fraction of the currently known Jovian and Neptunian Trojans (Fleming & Hamilton 2000; Chiang & Lithwick 2005), our results suggest that these populations could be comprised primarily of the survivors of those captured during the migration of the outer planets. Additionally, our results suggest that the dynamical structure and distribution of these clouds may contain key information that could shed light on the nature of migration during the latter stages of planet formation. That the dynamically excited Jovian Trojans could be linked to the dynamics of the early Solar system is reasonably well established – a number of authors (e.g. Morbidelli et al. 2005) have suggested that a complex chaotic orbital evolution followed by the migration of the outer planets could lead to such a distribution. However, because our simulations employed a pre-determined (non-chaotic) dynamical behaviour for the planets, one can conclude that the dynamically excited populations of Jovian and Neptunian Trojans can be taken only as evidence that these planets migrated from their formation sites, and not that their migration behaviour was chaotic during the early Solar system. Indeed, we stress that chaotic capture of Jovian Trojans operated well in our calculations without requiring Jupiter and Saturn to cross their mutual 1:2 MMR as advocated in the Nice model (Morbidelli et al. 2005; Levison et al. 2008 and references therein). This suggests that scenarios in which both giant planets start already locked in (as in the model of Thommes et al. 2008) or slightly outside this resonance, prior to planet migration, are equally capable of producing significant populations of captured Jovian Trojans. Therefore, it is plausible that captured Trojans obtained during a gentle migration of the outer planets might be sufficient to explain the currently known Jovian Trojan populations, so long as a sufficient amount of material were scattered to planet-crossing orbits during the period of planetary migration. In line with this idea, we have already shown that such a migration-capture scenario could explain the currently known Neptunian Trojans (Lykawka et al. 2009).

---

[6] These fractions are based on the results of Horner & Lykawka (2009) and Lykawka et al. (2009, in preparation). Nesvorny & Vokrouhlicky (2009) reported larger survival fractions at 1 Gyr (by a factor of several times). However, their values are obtained from runs not followed for the full 4 Gyr. Alternatively, the possible discrepancy with our obtained values might be explained by differences in the model parameters used, in particular those regarding the evolution of Neptune and the properties of the planetesimal disk.

On the other hand, an important caveat with the chaotic capture scenario is that the obtained captured Trojan populations do not appear to fully explain the observed distribution of Trojans in eccentricity and inclination space. For Jovian Trojans, the bottom panel of Fig. 1 reveals two regions in $e$-$i$ space that the model fails to populate, as evinced by both the lack of Trojans with $i > 30°$ and the absence of Trojans with $e \sim 0.1$-$0.15$ and $i < 10°$ (a conservative estimate of the breadth of a region which could easily be extended to $e \sim 0.1$-$0.2$ and $i < 15°$). Indeed, a close inspection of Fig. 1 strongly suggests that the lack of high inclination Trojans is the result of dynamical instability, as the majority of high-$i$ captured Trojans are lost over 4 Gyr, whilst the lack of intermediate eccentricity objects is not a result of the long-term evolution of the system, but rather reflects the primordial conditions after the capture mechanism ceased billions of years ago (compare both panels in Fig. 1). It is worth noting that collisional and long term dynamical effects cannot help to rectify the lack of objects in these regions, even acting over a number of Gyr, since the instability of the high $i$-region and difficulty to enter the low-$e$ regime remain valid even for "large" Trojans (Fig. 1). Furthermore, those Trojans found to be stable on Gyr-timescales did not suffer appreciable eccentricity or inclination changes over 4 Gyr. Intriguingly, the same problems also afflict the results of the Nice model, with the under-populated areas described above being clearly visible in both Fig. 1 of Levison et al. (2009) and Fig. 2 of Morbidelli et al. (2005) (indeed, the $e$-$i$ plot shown in that work is remarkably similar to the top panel of our Fig. 1).

For the Neptune Trojans, such a detailed comparison between the observed and calculated populations is not possible since only six Neptune Trojans have been discovered to date. Despite this, one potential problem with dynamical capture models is their failure to produce significant numbers of captured Trojans moving on highly inclined orbits ($i > 20°$) (Sheppard & Trujillo 2006; Nesvorny & Vokrouhlicky 2009). Our results on the long term evolution of both captured and transported Neptune Trojans (transported objects being ones which formed within the planet's 1:1 MMR, and were transported by it through its migration) seem to reconfirm this problem (Lykawka et al. 2009, in preparation). It is clear that future investigations will need to revisit the key parameters of the model in order to attempt to solve these problems for the captured Trojans on both planets. One possible solution (as discussed in Nesvorny & Vokrouhlicky (2009) for the Neptune Trojans) could be the use of initially excited (rather than dynamically cold) planetesimal disks.

Other avenues for future work include the determination of the true capture efficiency of Trojans by Saturn, and attempting to refine and improve our estimates of the total mass of primordial Neptunian Trojans. Over the coming years, our observational knowledge of the Neptunian Trojan population will grow rapidly, as an ever increasing number are detected by new surveys and improved astronomical instrumentation (such as Pan-STARRS (Jewitt 2003) and the LSST (Ivezic et al. 2008)). This will allow the calculation of the initial mass of the primordial Neptunian Trojan population, when combined with dynamical studies of the stability of the Neptune Trojans. These discoveries will provide a vital test-bed for studies of the formation and early evolution of the outer Solar system, and it is vital that models of planetary formation and migration attempt to predict the distributions of small bodies that could be observed, rather than just attempting to fit their results to the current state of play.

## 6 CONCLUSIONS

Overall, we can conclude that a gentle migration of the giant planets can lead to the perturbation of the primordial planetesimal disk in such a way that a small but significant fraction of bodies from that disk can be captured as Trojans by all of the giant outer planets. Such dynamical capture typically occurs as the planets themselves experience repeated crossings of their mutual mean-motion resonances, a process known as chaotic capture mechanism. A fraction of these captured bodies will then be expected to survive for long periods once the migration has come to a halt. Indeed, in the cases of Jupiter and Neptune, whose Trojan clouds are regions of well-documented high stability, it is plausible that such dynamical capture may explain the entire vast population of Trojans believed to be hosted by these planets. This opens a new window for the investigation of the origin of planetary Trojans, and adds weight to the idea that the distribution of Trojans in orbital element space could yield a wealth of information on the conditions in which the planets first formed and then evolved (planetary migration). Finally, it should be noted that the trans-Neptunian disk would not have been the only source of planet-crossing objects during the period of planetary migration. In addition to the debris remaining in the vicinity of the planets, from which they formed, is it likely that other reservoirs (such as the outer asteroid belt) would have been perturbed and disrupted, leading to a significant additional flux of material to planet-encountering orbits. Such material will have no doubt swelled the burgeoning Trojan populations, and it is likely that the modern Trojan clouds contain objects which formed at a wide variety of locations throughout the proto-planetary nebula, acting as effective archives detailing the conditions from which our planetary system formed.

## ACKNOWLEDGEMENTS
We would like to thank an anonymous referee for a number of helpful comments and suggestions, which allowed us to improve the overall presentation and flow of this work. PSL and JAH gratefully acknowledge financial support awarded by the Daiwa Anglo-Japanese Foundation and the Sasakawa Foundation, which proved vital in arranging an extended research visit by JAH to Kobe University. PSL appreciates the support of the COE program and the JSPS Fellowship, while JAH appreciates the support of STFC.

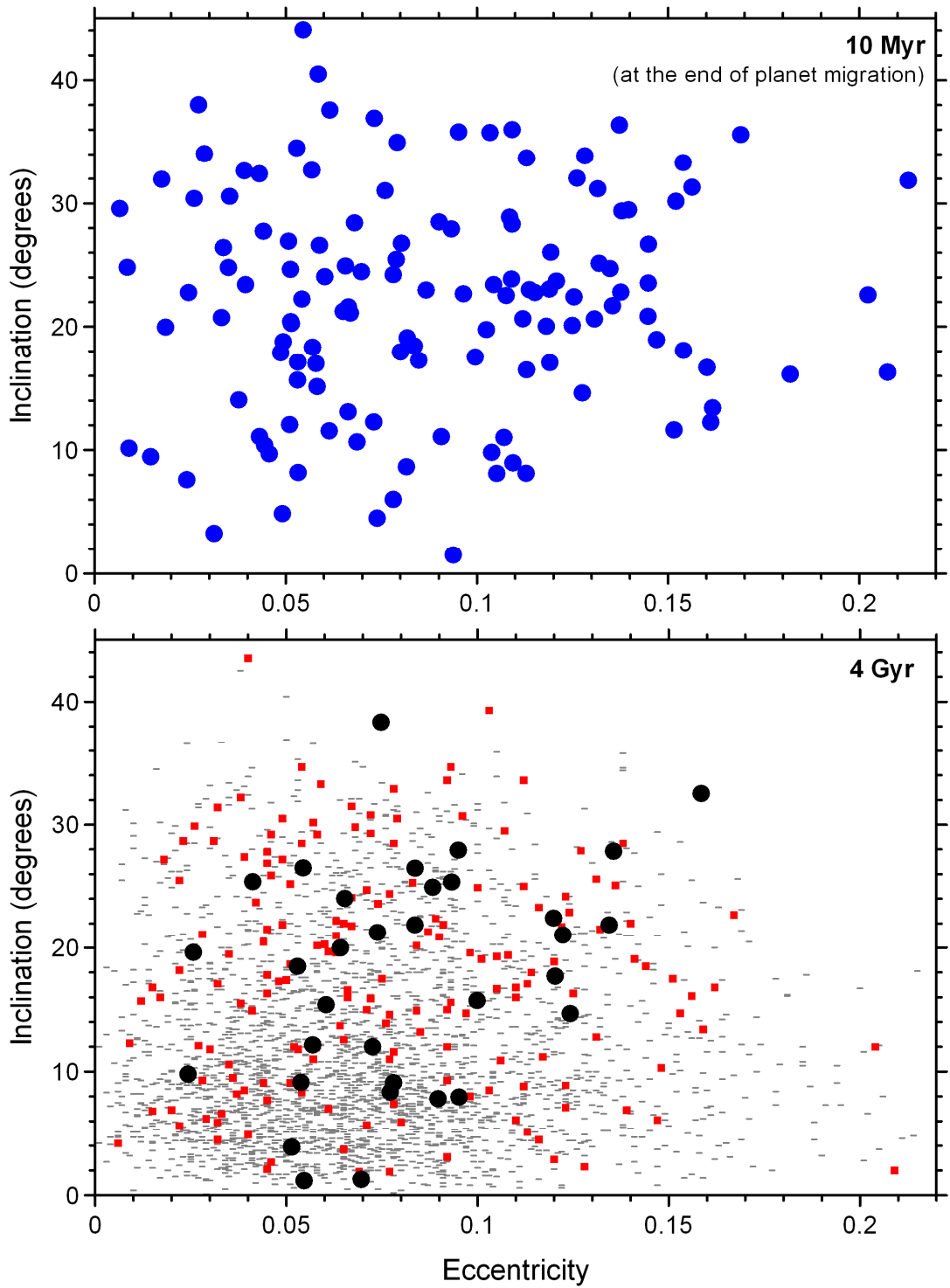

**Figure 1:** Plot showing the orbital distribution of objects captured as Jovian Trojans once the migration of the giant planets had come to an end (top), and after following the evolution of the same system for a period of 4 Gyr (bottom). The objects plotted were all found to have been moving on tadpole orbits around the L4 and L5 Jovian Lagrange points. Currently known Trojans with orbits with at least two opposition observations are shown for comparison in the bottom panel, taken from the IAU Minor Planet Center[*] on 25$^{th}$ November, 2009. In this panel, "large" Trojans (sizes > 50 km) with absolute magnitudes, *H*, less than 10.5 are represented by squares, whilst "small" Trojans (sizes < 50 km; *H* > 10.5) are shown as minus signs.

---

[*] http://www.cfa.harvard.edu/iau/lists/JupiterTrojans.html

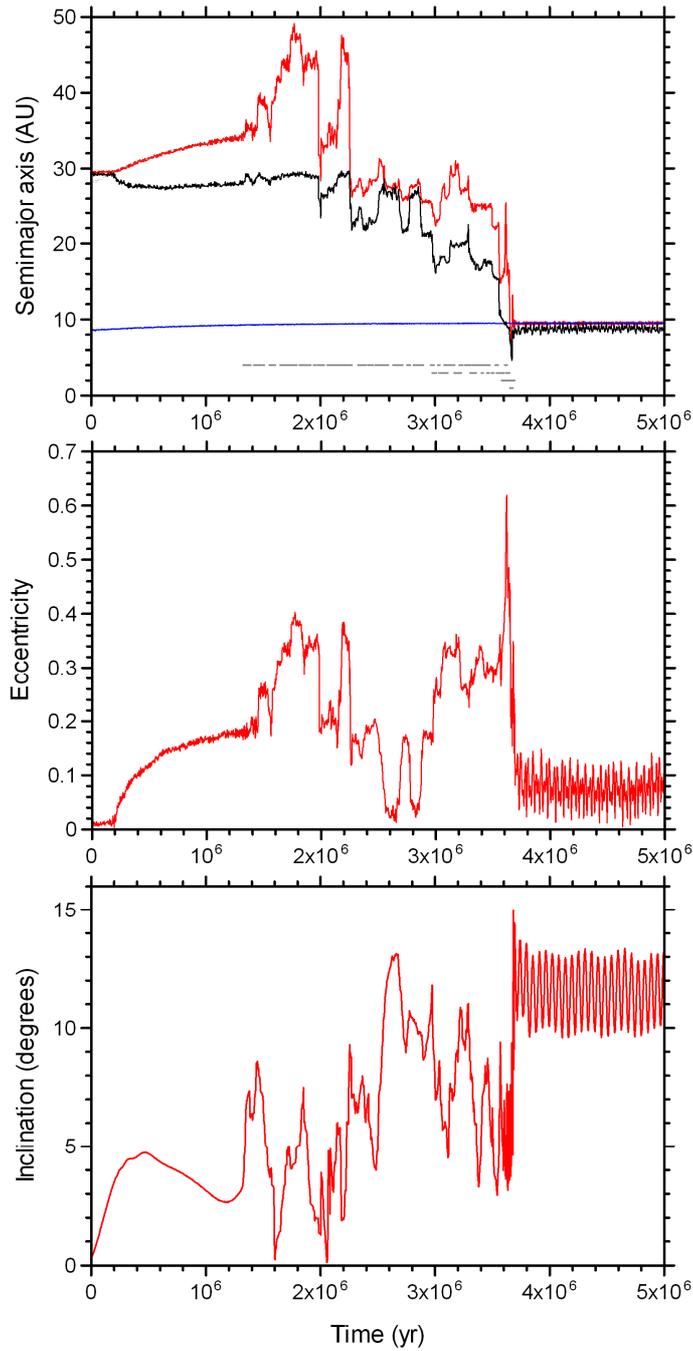

**Figure 2:** An exemplar case of a disk planetesimal being captured in the Saturnian Trojan cloud during a simulation of the rapid migration of Neptune from 23.1 AU to its current location. The plots detail the evolution of the objects semi-major axis (top, red line), eccentricity (middle) and inclination (bottom) over the total integration time for the simulation (a period of 5 Myr). The object's perihelion distance (plotted in black) and the evolution of Saturn's semi-major axis (in blue) are also shown in the upper panel. The object was trapped in the 4:3 MMR with Neptune between 0.2 and 1.3 Myr after the start of the integration, after which it left the resonance and started to experience close encounters with Neptune. Later, the object underwent a period of complex orbital evolution resulting from close encounters with the other giant planets. The timeline of such close encounters is shown at the bottom of the upper panel, where gray squares indicate close encounters between the particle and a given giant planet. The upper row of squares shows encounters with Neptune, the next with Uranus, then Saturn and finally Jupiter (lowest). Finally, the object becomes captured as a stable Saturnian Trojan after ~3.7 Myr, and remains such until the end of the simulation.

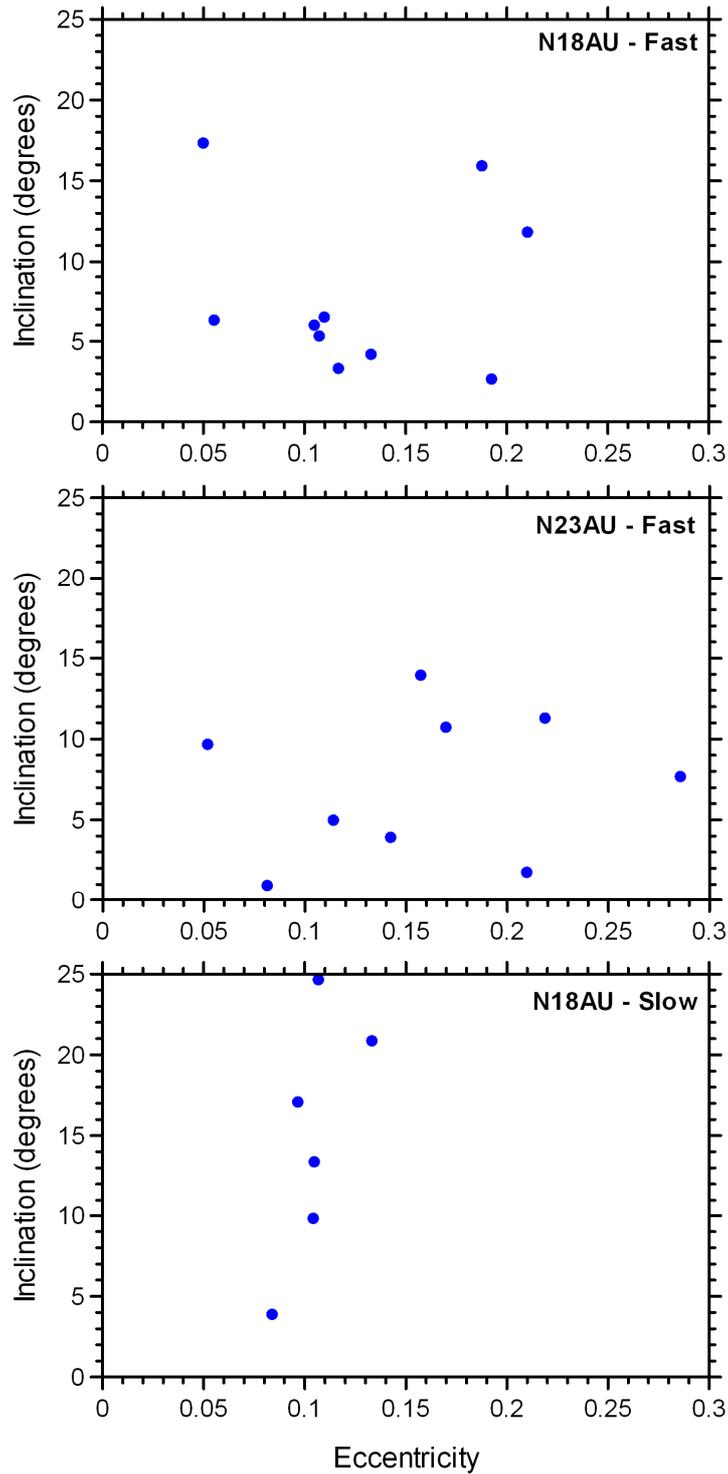

**Figure 3:** Plot showing the orbital distribution of objects captured as Uranian Trojans once the migration of the giant planets had come to and end. The three frames show the captured objects for each of the three scenarios that were found to result in Uranian Trojans. From top to bottom, the panels show the following case. Top: Uranus and Neptune migrated rapidly to their current locations from initial distances of 14.1 and 18.1 AU respectively. Middle: Uranus and Neptune migrated rapidly to their current locations from 16.1 and 23.1 AU respectively. Bottom: Uranus and Neptune migrated slowly to their current locations from initial distances of 14.6 and 18.1 AU, respectively. The objects plotted were found to have been moving on tadpole orbits around the L4 and L5 Uranian Lagrange points for at least 0.5 Myr and 1 Myr by the end of the simulations detail fast and slow migration scenarios, respectively.

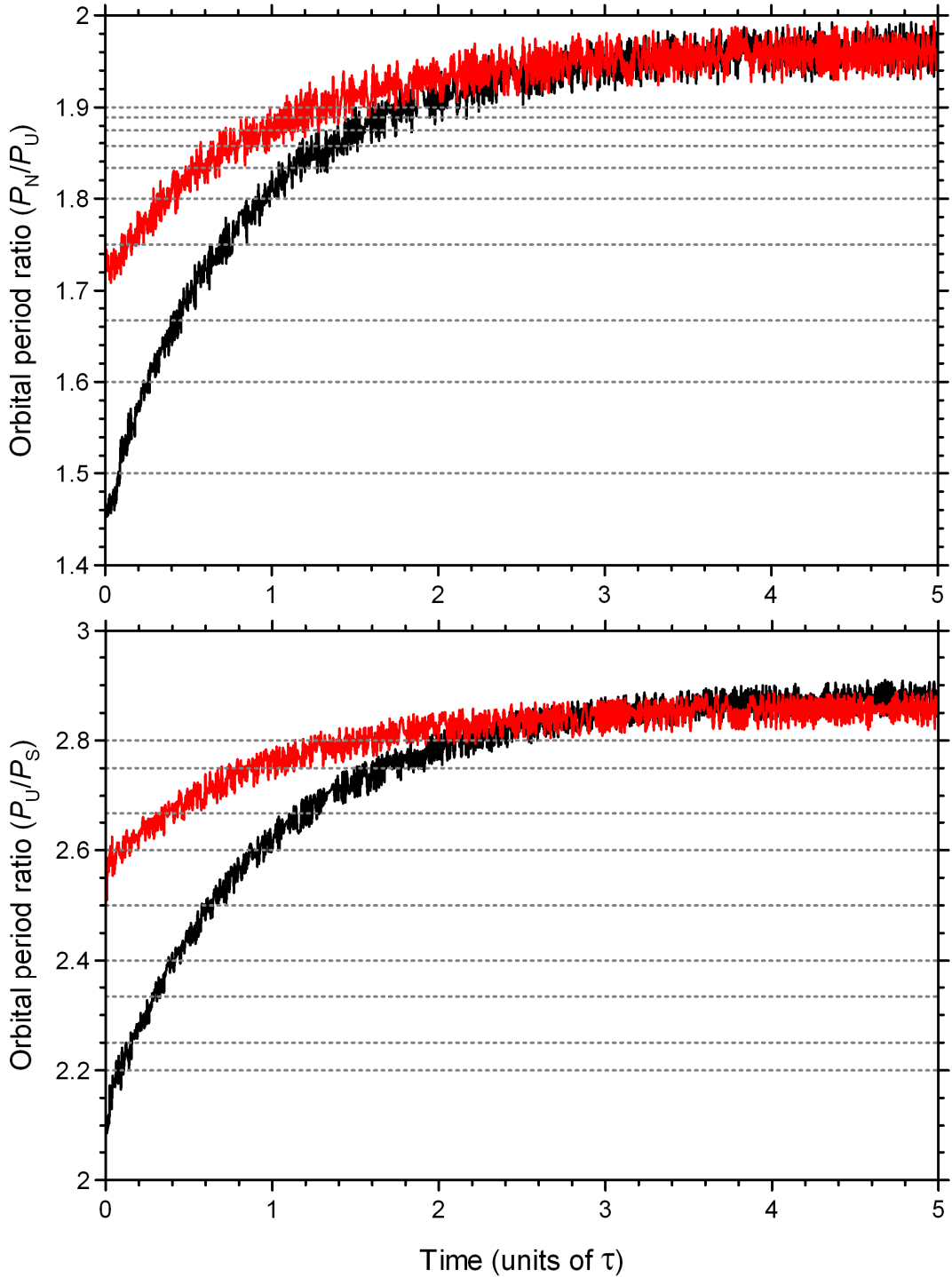

**Figure 4:** The evolution of the orbital period ratios between Neptune and Uranus ($P_N/P_U$) (top) and Uranus and Saturn ($P_U/P_S$) (bottom) as a function of time during the fast migration runs. The evolution of these quantities during the N18-F and N23-F simulations are plotted as black and red curves, respectively. The elapsed integration time is given in units of the migration timescale, $\tau$, as described in the text. The evolution for slow migration runs was very similar, as a result of similar initial conditions (Table 1). Saturn, Uranus and Neptune clearly experienced several mutual MMR crossings during their migration. Top: Horizontal lines represent examples of mutual MMRs between Uranus and Neptune, namely 2:3, 5:8, 3:5, 4:7, 5:9, 6:11, 7:13, 8:15, 9:17, and 10:19 (in order of increasing $P_N/P_U$). Bottom: Horizontal lines represent examples of mutual MMRs between Saturn and Uranus, namely 5:11, 4:9, 3:7, 5:12, 2:5, 5:13, 3:8, 4:11, and 5:14 (in order of increasing $P_U/P_S$).